\documentclass[%
 preprint, 
superscriptaddress,
 amsmath,amssymb,
 aps, physrev,
]{revtex4-2}

\usepackage{graphicx}
\usepackage{dcolumn}
\usepackage{bm}

\begin{document}

\preprint{APS/123-QED}

\title[Multi-photon enhanced resolution for Superconducting Nanowire Single-Photon Detector-based Time-of-Flight lidar systems]{Multi-photon enhanced resolution for Superconducting Nanowire Single-Photon Detector-based Time-of-Flight lidar systems}

\author{Adrian S. Abazi}
\affiliation{Department for Quantum Technology, University of Münster, Heisenberstr. 11, 48149 Münster, Germany}%
\affiliation{Center for Nanotechnology (CeNTech), Heisenbergstr. 11, 48149 Münster, Germany}%
\affiliation{Center for Soft Nanoscience (SoN), Busso-Peus-Str. 10, 48149 Münster, Germany}

\author{Roland Jaha}%
\affiliation{Center for Nanotechnology (CeNTech), Heisenbergstr. 11, 48149 Münster, Germany}%
\affiliation{Center for Soft Nanoscience (SoN), Busso-Peus-Str. 10, 48149 Münster, Germany}
\affiliation{Kirchhoff-Institute for Physics, Heidelberg University, Im Neuenheimer Feld 227, 69120 Heidelberg, Germany}%

\author{Connor A. Graham-Scott}
\affiliation{Department for Quantum Technology, University of Münster, Heisenberstr. 11, 48149 Münster, Germany}%
\affiliation{Center for Nanotechnology (CeNTech), Heisenbergstr. 11, 48149 Münster, Germany}%
\affiliation{Center for Soft Nanoscience (SoN), Busso-Peus-Str. 10, 48149 Münster, Germany}

\author{Wolfram H. P. Pernice}
\affiliation{Kirchhoff-Institute for Physics, Heidelberg University, Im Neuenheimer Feld 227, 69120 Heidelberg, Germany}%

\author{Carsten Schuck}
 \email{Corresponding author: carsten.schuck@uni-muenster.de}
\affiliation{Department for Quantum Technology, University of Münster, Heisenberstr. 11, 48149 Münster, Germany}%
\affiliation{Center for Nanotechnology (CeNTech), Heisenbergstr. 11, 48149 Münster, Germany}%
\affiliation{Center for Soft Nanoscience (SoN), Busso-Peus-Str. 10, 48149 Münster, Germany}

\date{\today}

\begin{abstract}
Superconducting nanowire single photon detectors (SNSPDs) emerged in the last decade as a disruptive technology that features performance characteristics, such as high sensitivity, dynamic range and temporal accuracy, which are ideally suited for light detection and ranging (lidar) applications. Here, we report a time-of-flight (TOF) lidar system based on waveguide-integrated SNSPDs that excels in temporal accuracy, which translates into high range resolution. For single-shot measurements, we find resolution in the millimeter regime, resulting from the jitter of the time-of-flight signal of 21$\,$ps for low photon numbers. We further decrease this signal jitter to 11$\,$ps by driving the SNSPD into a multiphoton detection regime, utilizing laser pulses of higher intensity, thus improving range resolution. For multi-shot measurements we find sub-millimeter range-accuracy of 0.75$\,$mm  and reveal additional surface information of scanned objects by visualizing the number of reflected photons and their temporal spread with the acquired range data in a combined representation. Our realization of a lidar receiver exploits favorable timing accuracy of waveguide-integrated SNSPDs and extends their operation to the multiphoton regime, which benefits a wide range of remote sensing applications. 
\end{abstract}

\maketitle

\section{\label{sec:intro} Introduction}
Light detection and ranging (lidar) is a remote sensing technique based on sending and receiving photons reflected off a distant object. One obtains the distance to the object by measuring the time-of-flight (TOF) of photons traveling to it and back, either by direct detection using a pulsed or amplitude modulated laser or via coherent detection and interference using a continuous-wave laser modulated in frequency. Precisely controlled laser sources and detectors with high timing accuracy thus make lidar systems a key technology in remote sensing systems for robotics and vehicles \cite{Royo2019,Roriz2022}, atmospheric \cite{Wulfmeyer2015,Heinze2017,Ehret2017,Comeron2017}, geological \cite{Lim2003,Hudak2009} or agricultural \cite{Debnath2023} observations as well as the generation of digital twins of real-world objects \cite{Xue2020,Brock2021}.
State-of-the-art lidar systems further reach sensitivities down to the single-photon level, enabling applications under challenging conditions, such as determining ranges of thousands of kilometers with centimeter-level precision in space applications \cite{Li2016,Xue2016}, 3D imaging over km distances \cite{McCarthy2013,Li2021,McCarthy2025} or through partially transparent or scattering media \cite{Tobin2019,Wallace2020} and non-line-of-sight 3D imaging \cite{Liu2019} on shorter distances.
This sensitivity additionally benefits the acquisition speed of photon counting lidar, as 3D imaging is  possible by detecting the first photon only \cite{Kirmani2014}, minimizing the integration time per pixel to a single time interval or single-shot of the pulsed laser. This effect applies for direct detection lidar, since coherent detection lidar intrinsically requires multiple detections. For direct detection lidar with the fastest acquisitions speeds, i.e. range information from only a few or a single-photon, the range precision predominantly depends on the temporal resolution of the receiver system. The dominating contribution for all ranges under space conditions and for shorter ranges under atmospheric conditions is the detector jitter. For longer ranges under atmospheric conditions, the detector jitter influences the integration time necessary to reach maximum precision. Here, we show that low-jitter superconducting nanowire single-photon detectors (SNSPDs) improve the precision of photon counting lidar systems for single-shot measurements and illustrate how to reveal additional information about remote objects from multi-shot measurements, often referred to as "full waveform" lidar data \cite{Wallace2020}.\\

A lidar system fundamentally consists of a transmitter and receiver unit. Here we will focus on the single photon detector as the most relevant component of a photon counting lidar system. While requirements and state-of-the-art vary between different applications \cite{Behroozpour2017,Hadfield2023,Guan2022,Li2022}, several detector performance parameters need to be optimized. Firstly, high sensitivity will benefit the signal-to-noise ratio (SNR) that can vary strongly due to environmental influences. Secondly, a high dynamic range allows for recording large variations in the measured signal due to its strong dependence on distance and reflectivity of the scanned object as well as other environmental factors. Thirdly, high timing accuracy directly translates to improved range resolution. Lastly, a wide spectral performance is desirable, to allow for differing spectral reflectivity and absorption properties of targets, to account for bandwidth requirements resulting from the type of laser source \cite{Li2022}, and to enable applications based on multi-spectral and spectroscopic methods \cite{Comeron2017,Li2021a}.\\
SNSPDs are a popular choice for high-performance single-photon detectors in quantum technologies \cite{Natarajan2012,Wang2019,Moody2022}, because they offer very high sensitivities through combining near unity detection efficiency \cite{Pernice2012,Ferrari2018,Wolff2020} with very low dark count rates (DCR) even reaching the millihertz regime \cite{Schuck2013}. Moreover, SNSPDs provide a high dynamic range \cite{Tiedau2019}, allowing for high maximum count rates in the GHz range \cite{Craiciu:23}. Compared to single-photon avalanche diodes (SPADs) \cite{Bugge2014}, they possess a more robust detection mechanism, because the current through the photosensitive element is massively suppressed during the detection, with the device switching to a transient resistive state.
Importantly, SNSPDs also offer high timing accuracy due to low jitter of the signal latency, typically in the range of tens of ps, which has already been exploited in ranging applications, such as single-photon lidar \cite{McCarthy2013,Li2016,Taylor2020,McCarthy2025} and optical time domain reflectometry \cite{Schuck2013a}. The lowest jitter reported for SNSPDs is below 3$\,$ps \cite{Korzh2020} and had been exploited for laser ranging with millimeter-scale depth resolution using tens of detected photons, however with minimal absorption efficiency of the active detector element. 
Lastly, SNSPDs also satisfy the spectral bandwidths requirements of lidar systems, offering sensitivity from the ultraviolet to the mid-infrared wavelength range \cite{Marsili2012,Wolff2021,Taylor2023}.\\

Conventional SNSPDs typically operate under direct illumination of a meandering superconducting nanowire from an optical fiber, resulting in comparatively large on-chip footprints with corresponding timing characteristics. Instead, we here employ waveguide integrated (wi-)SNSPDs. Compared to irradiance of a nanowire from an optical fiber under normal incidence, the photons traveling inside a waveguide are absorbed along their direction of propagation. This allows for efficient detection with significantly shorter nanowires, which exhibit lower temporal jitter values \cite{Korzh2020}. Here, we exploit this favorable combination of low jitter and efficient photon counting in wi-SNSPDs for a TOF-based lidar system. Moreover, we show how to reduce the temporal jitter of the SNSPD by driving it into a multiphoton detection regime, using higher intensity laser pulses for ranging. We exploit these features for acquiring high-resolution lidar images of remote objects, thus enabling a wide range of ranging applications, such as the generation of accurate digital twins of real-world objects. 

\section{\label{sec:methods} Photon counting Time-of-Flight lidar}
In a direct detection TOF lidar scheme, the time-of-flight or travel time, $ t_{TOF}$, of a short laser pulse reflected off a remote object yields the distance to the object
\begin{equation}
    \label{eqn:TOF_distance}
    d = \frac{c}{2} ( t_{TOF}-t_0 ),
\end{equation}
where $c$ is the speed of light and $t_0$ denotes a temporal offset corresponding to the origin of the range scale, i.e. the time when the laser pulse is launched from the aperture of the ranging system. The measurement uncertainty of the distance $\Delta d = \frac{c}{2}\cdot \Delta t_{TOF}$ is directly dependent on the timing accuracy of all signals that contribute to the latency and introduce a statistical spread of the resulting TOF-signal latency, i.e. jitter, 
\begin{equation}
    \label{eqn:jitter_contrib}
    \Delta t_{TOF} = \sqrt{\Delta t_{SNSPD}^2 + \Delta t_{l}^2 + \Delta t_{TDC}^2+ \Delta t_{sync}^2 + \Delta t_{r}^2}.
\end{equation}
We here identify relevant contributions due to the jitter of the SNSPD output signal, $\Delta t_{SNSPD}$, the duration of the laser pulse, $\Delta t_{l}$, the jitter of the time to digital conversion (TDC) system, $\Delta t_{TDC}$, the jitter of the synchronization signal, $\Delta t_{sync}$, and contributions due to pulse broadening during propagation and reflection off the target, $\Delta t_{r}$. Because the start signal for measuring the time-of-flight, $ t_{TOF}$, is triggered by the laser pulse itself, temporal fluctuations when the laser pulse is emitted, i.e. the jitter of the laser, do not contribute to $\Delta t_{TOF}$.

\subsection{\label{sec:methods_nanoFab} Superconducting Nanowire Detectors for Lidar} 
Designing the detector for a high-precision lidar system, requires the reconciliation of tradeoffs between the temporal jitter, maximum count rates, detection efficiency and noise. 
For superconducting nanowire detectors, the jitter characteristics depend on intrinsic superconducting fluctuations for a given material, nanowire geometry and the electrical readout circuit \cite{Sidorova2017,Ferrari2018,Korzh2020}. We here consider two relevant SNSPD-designs that address trade-offs between detector-jitter, detection efficiency and signal-to-noise performance in conceptually different ways. Design SNSPD1 is optimized for low jitter performance sacrificing detection efficiency, while design SNSPD2 optimizes on-chip detection efficiency at the expense of somewhat increased jitter values. \\

Our designs are guided by considering that the jitter of the detector signal is dependent on Fano fluctuations \cite{Kozorezov2017} of the energy deposited in the electronic system, even if the absorbed photons are monochromatic. This dependence is strongly influenced by the bias current, which appears to flatten the slope of any latency dependence, leading to lower jitters for higher bias currents \cite{Allmaras2019,Korzh2020}. The achievable bias currents and the shape of the latency dependence are, in turn, determined by the composition and quality of the superconducting material, along with phonon dispersion, which is influenced by the surrounding materials and temperature. \\

Moreover, the jitter performance of a nanowire detector is influenced by its kinetic inductance $L_K$, geometry and electronic noise. Due to the wire's kinetic inductance, electrical signals travel slower than the speed of light and fluctuations in the locations where photons are absorbed then increase such geometrical jitter $j_{geom}$ \cite{Calandri2016}. Inhomogeneities of the geometry along the wire, resulting in inhomogeneity of $L_K$ also contribute to the jitter. Importantly, electronic noise $\delta V$ of the system causes temporal fluctuations, when the voltage response of the detector crosses the threshold at the time-to-digital converter, resulting in an electronic jitter component $j_{electro}$. This time when crossing the voltage trigger level is a fraction of the full risetime $\tau_{rise}$ of the voltage response, which is also dependent on the kinetic inductance \cite{Nicolich2019}, resulting in the dependence, 
\begin{equation}
j_{electro} \propto \delta V \tau_{rise} \propto \delta V \sqrt{L_K},
\end{equation} for the electronic jitter \cite{Jaha2024}. The kinetic inductance $L_K \propto \frac{l}{A}$ itself scales geometrically with the length $l$ and the inverse of the cross-sectional nanowire area $1/A$.
Ideally, one hence reduces the length and increases the cross-section for achieving lower kinetic inductances $L_K$ and therewith lower jitter.  However, the cross-section cannot be increased indefinitely because the quantum efficiency decreases for larger cross-sections, once it exceeds a critical dimension defined by the photon energy \cite{Semenov2005,Hofherr2010}. Moreover, reducing the nanowire length reduces its absorption efficiency. \\

\begin{figure}
\includegraphics{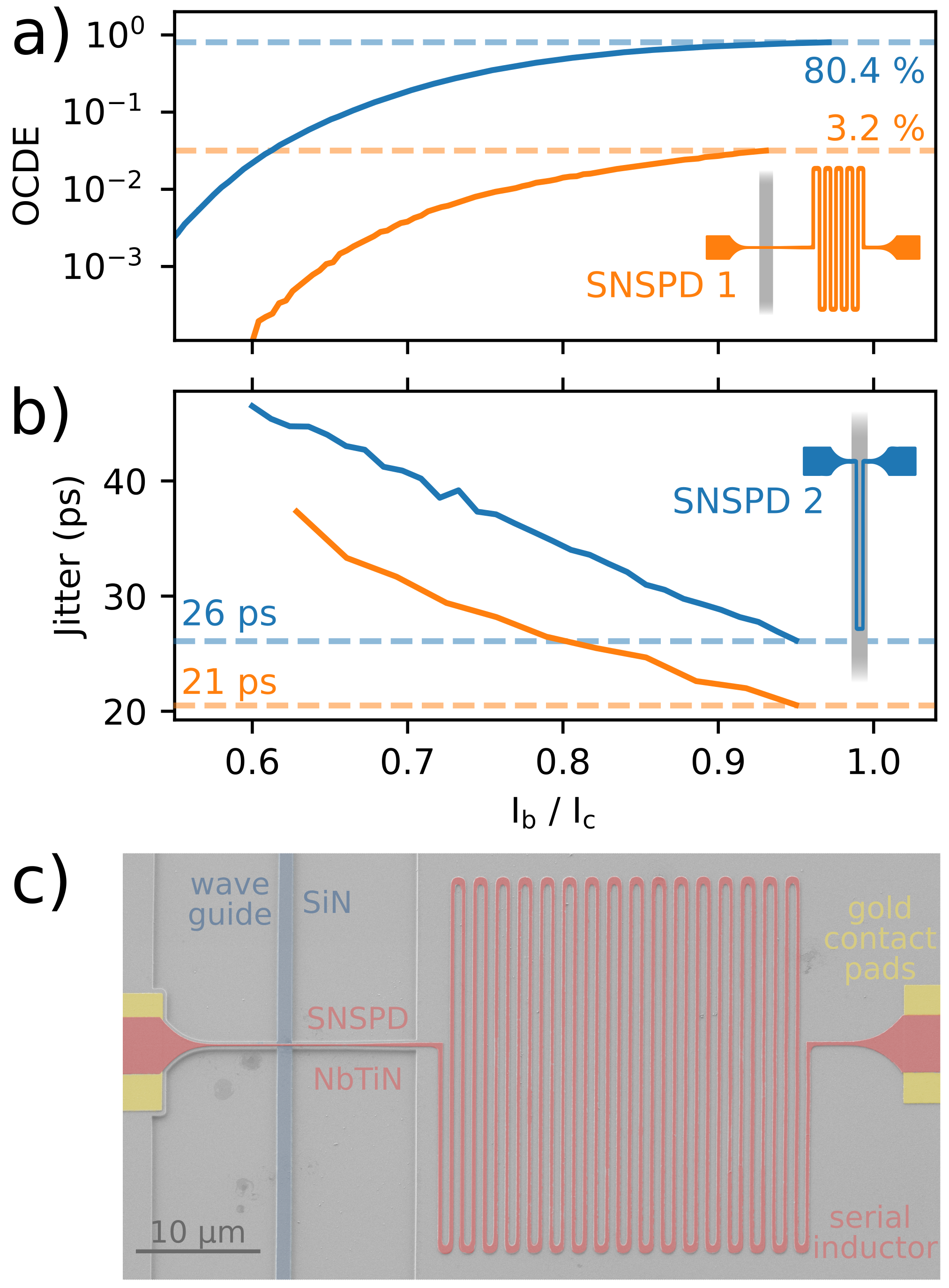}
\caption{\label{fig:snspds}
Performance comparison of two SNSPD designs, SNSPD1 being a low kinetic inductance wire (thick and short) and SNSPD2 being a high kinetic inductance wire (thin and long), for varying bias currents $I_b$ relative to the switching currents $I_{c1} = 23.0\, \mathrm{\mu A}$ and $I_{c2} = 10.6\, \mu A$. a) depicts the on-chip detection efficiency (OCDE) and b) the timing jitter.  c) False-colored scanning electron micrograph depicting SNSPD1 (red) including the SiN-waveguide (blue), and electrical contact pads (yellow).
}
\end{figure}

As a result of these considerations, we chose to minimize the kinetic inductance for SNSPD1, by fabricating a nanowire with large cross-section (12x100)$\,$nm$^2$ and short length (10$\,$µm) from NbTiN. A horizontal orientation to the photonic waveguide, as depicted schematically in Figure \ref{fig:snspds} a), reduces the interaction width to the waveguide width of 1.2$\,$µm, thus minimizing the geometrical jitter. As such low inductance nanowires have a large probability to latch, i.e. to switch into a permanent resistive state, we include an inductor (300$\,$nm width x 800$\,$µm length) in series with the detecting nanowire to allow for higher biasing currents.
For SNSPD2, on the other hand, we aim for higher detection efficiency, utilizing a (5x120)$\,$nm$^2$ cross-section and a nanowire length of 50$\,$µm, here made from NbN. The arrangement in a "U-" or "hairpin-" shape on top of the waveguide with three bends as shown in Figure \ref{fig:snspds} b) enables efficient absorption of photons traveling inside the waveguide but adds kinetic inductance and geometric jitter.
Figure \ref{fig:snspds} a) depicts the on-chip detection efficiency (OCDE) of the two SNSPDs in relation to their bias currents $I_b$.
In accordance with their geometric design, SNSPD1 indeed shows a significantly higher detection efficiency of photons travelling in the waveguide of 80.4\% compared to SNSPD2, which achieves 3.2\% at the highest bias current before latching. 
Fig.$\,$\ref{fig:snspds} b) depicts the jitter of the SNSPDs for varying $I_b$. With increasing bias current, $I_b$, the jitter decreases. At bias currents of 95\% of the switching current we find that SNSPD1 shows a lower jitter performance of 21$\,$ps as compared to 26$\,$ps for SNSPD2, which reflects the higher switching current of 23.0$\,\mathrm{\mu A}$ for SNSPD1 as compared to 10.6$\,\mathrm{\mu A}$ for SNSPD2.
We observe the onset of saturation behavior for the OCDE with an increase in the bias current, indicating that even lower jitter performance may be possible for optimized fabrication of the superconducting nanowires.\\
We fabricate SNSPDs in a top-down approach utilizing magnetron sputtering of 12$\,$nm NbTiN (SNSPD1) or 5$\,$nm NbN (SNSPD2) on 330$\,$nm silicon nitride (SiN) on insulator samples. Nanowires and nanophotonic waveguides are patterned in 100kV electron-beam lithography and fluorine-based reactive ion etching \cite{Beutel2022,Wolff2021,Haeussler2023}. An exemplary false colored scanning electron micrograph of a fabricated device, here SNSPD1, is depicted in Fig.$\,$\ref{fig:snspds} c)\\
 
\subsection{\label{sec:methods_setup} Experimental Setup and Data Processing}
\begin{figure*}
\includegraphics{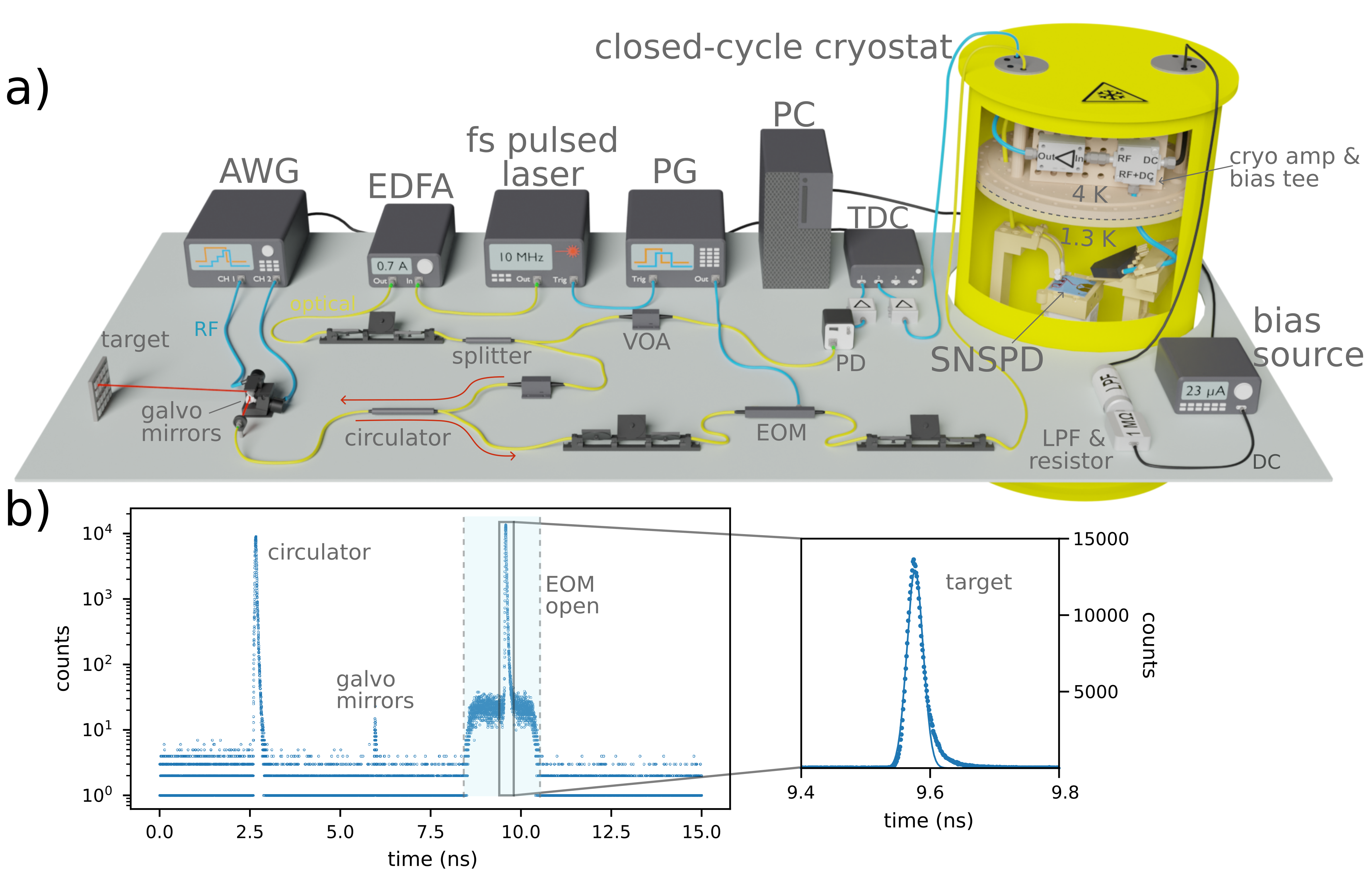}
\caption{\label{fig:setup}a) Setup of the SNSPD-based TOF lidar system. Laser pulses are reflected off a target (left), collected, and detected with an SNSPD. The time of flight is determined via the latency with respect to a reference signal on a photo-diode (PD) and a time-to-digital converter (TDC), the galvo mirrors scan the beam over the target, creating an image by repeating the TOF-measurement for varying angles. b) Exemplary histogram of a single pixel TOF-measurement, depicting reflected counts vs. latency time differences.}
\end{figure*} 
Figure \ref{fig:setup} a) shows the experimental setup of the SNSPD-based TOF lidar system. We cool the nanowire devices to 1.3$\,$K in a closed-cycle cryostat (ICE Oxford) and bias them near the switching current, utilizing a bias-T with a low pass filter (LPF), a 1$\,\mathrm{M\Omega}$ load resistor and a current source on the DC port, as well as two low noise amplifiers on the RF port for the signal read out. Utilizing a cryogenic amplifier and a high accuracy time-to-digital-converter (TDC, TimeTagger X, Swabian Instruments) minimizes the system contributions to the overall jitter, $\Delta t_{TOF}$. 
We employ a 100$\,$fs pulsed laser system (Elmo, Menlo Systems) operating at wavelengths of 1560$\pm$30$\,$nm, which minimizes $\Delta t_{l}$. An erbium doped fiber amplifier (EDFA, Pritel, LHNPFA-30), increases the power of the laser pulses. We split off a fraction of the laser pulses for generating synchronization signals with a fast, 25 GHz, photo-diode receiver (PD, Newport, Model-1414) that provides a timing reference as input to the TDC. The remaining laser pulses are used for ranging, where a variable optical attenuator (VOA) allows for freely varying the laser pulse power. After passing a circulator (-60$\,$dB crosstalk), a fixed focus fiber collimator (2$\,$mm beam spot size) and a galvo mirror provide an interface to free-space propagation for scanning target objects. After the reflection off the target, back-scattered light is collected by the same fiber collimator and guided to the SNSPD via the circulator and an electro-optical modulator (EOM) for improving the ratio between the lidar signal and noise from the EDFA.\\
We record the latency time $t_{tof}$ between the start signal from the PD and the detection events at the SNSPD as multiple stop events in histograms, with an exemplary measurement shown in Fig.$\,$\ref{fig:setup} b). The digital bin width of the TDC is set to 1$\,$ps with a full width half maximum (FWHM) accuracy of 1.5$\,$ps per channel, minimizing the contribution to $\Delta t_{TDC}$. We find multiple peaks, each corresponding to reflections in the optical path. Additionally, the optical gating effect of the EOM is apparent with a $\approx$10$\,$db improvement of the SNR enabling operation of the system at higher laser powers before the SNSPD latches.\\
We fit Gaussian distributions to the signal that we associate with the target, centered at $t_{TOF}$, as shown in the inset of Fig.$\,$\ref{fig:setup} b). For scanning a target with the lidar setup, the two mirrors on the galvo-motor assembly are rotated by applying voltage with an arbitrary waveform generator (AWG). Histograms are recorded for each voltage/angle combination for an integration time of 100$\,$ms at the 10$\,$MHz repetition rate of the pulsed laser.
\section{\label{sec:results} Results}
\subsection{\label{sec:results_resolutions} Range resolution}
The spatial resolution of our lidar system in the lateral directions is determined by the beam spot size for close distances and the minimal rotation steps of the galvo mirrors for large distances. The axial depth or range resolution is related to the temporal resolution via equation (\ref{eqn:TOF_distance}). We define two regimes for this temporal resolution: "single-shot" measurements, corresponding to a single detection event per scanned pixel, and "multi-shot" measurements, involving statistical analysis of multiple detection events per pixel. Multi-shot measurements, also referred to as "full waveform" measurements \cite{Wallace2020}, hence sample the intensity of the returning wave and map the statistical spread of single-shot measurements. The single-shot regime allows for the fastest acquisition speeds, while the multi-shot regime offers intensity information and more precise temporal information at slower acquisition speeds.
\begin{figure}
\includegraphics{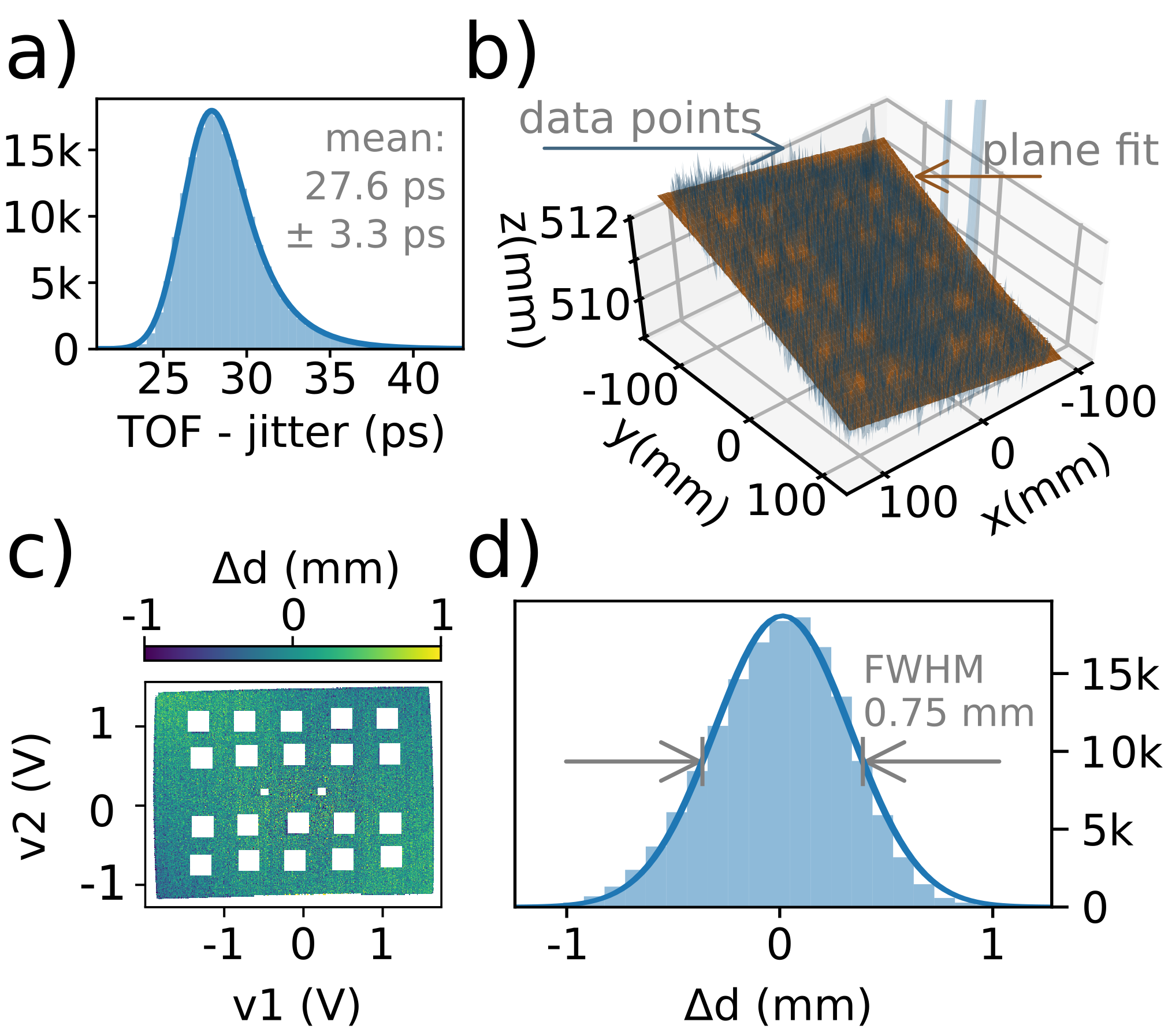}
\caption{\label{fig:acc}Determination of range resolution for single- and multi-shot measurements, using a calibration plate as a target, non-flat sections are removed from the measurement. a) shows a histogram of TOF-jitters, with an exponentially modified Gaussian fit, corresponding to the temporal single-shot resolution, b) measurement data in 3D space, with a plane fitted in orange, c) difference in distance $\Delta d$ of the measurement points to the fitted plane in dependence on the mirror voltages (excluded non-flat areas appear in white) and d) the distribution of the $\Delta d$ with a Gaussian fit, FWHM corresponding to the multi-shot measurement}
\end{figure}
To assess system performance, we scan a (200$\,$x$\,$250)$\,$mm$^2$ calibration plate containing flat and non-flat areas at a distance of 510$\,$mm from the last mirror of our lidar setup. We here employ SNSPD1 biased with 20$\,\mathrm{\mu A}$ current and record reflected photons with an integration time of 100$\,$ms. The distribution of the TOF-jitter values is depicted in Fig.$\,$\ref{fig:acc} a), considering the data of 186,000 measurements or pixels performed on only the flat areas of the calibration plate. We fit an exponentially modified Gaussian function to the distribution of the number of occurrences on the y-axis of Fig.$\,$\ref{fig:acc} a) and the TOF-jitter divided into intervals of 0.5$\,$ps on the x-axis. We find a mean jitter value of 27.6$\,$ps corresponding to a single-shot range resolution of 4.1$\,$mm.\\
Considering the contributions of the TOF-jitter, described in equation (\ref{eqn:jitter_contrib}), we can determine the contribution due to pulse broadening and reflection off the target, $\Delta t_r$, by comparing the TOF-jitter to the jitter of the SNSPD system shown in Fig.$\,$\ref{fig:snspds} b).
For the biasing conditions during TOF-data acquisition we find a corresponding jitter of the SNSPD system of 23.7$\,$ps, yielding a value of $\Delta t_r = 14.1\,$ps that results from the temporal fluctuations due to broadening of the laser pulse during propagation through air, fiber optic cables, the EOM and the circulator, as well as contributions from the reflection at the target itself.
Using the multi-shot $t_{TOF}$ data collected across the target area, which we assume to be flat, we fit a plane representing the target surface, as visualized in Fig.$\,$\ref{fig:acc} b). In Fig.$\,$\ref{fig:acc} c) and Fig.$\,$\ref{fig:acc} d) we visualize the distance variations, $\Delta d$, between the plane and the measured distances for the voltages of the two involved galvo mirrors and as a distribution of the occurrences, respectively. The variations $\Delta d$ range from -1$\,$mm to +1$\,$mm with the largest deviations appearing at the edges of the device, due to trapezoidal distortions resulting from the actuation of the galvo mirror. The range resolution of the multi-shot measurement thus corresponds to the 0.75$\,$mm FWHM of the distribution.

\subsection{\label{sec:results_image} Imaging, Data Analysis}
\begin{figure*}
\includegraphics{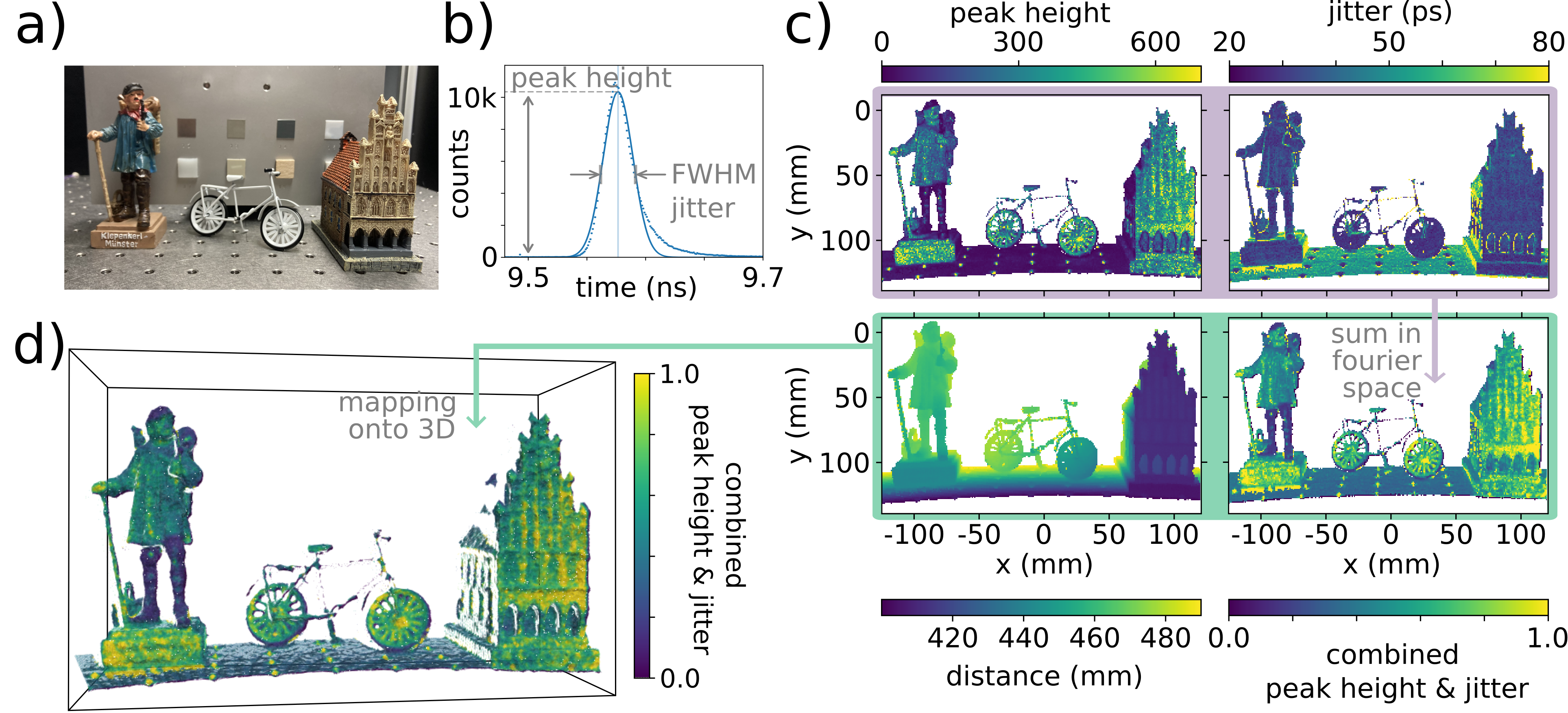}
\caption{\label{fig:image} Imaging example of the SNSPD-based lidar and data fusion technique, a) photograph of the scanned object, b) TOF-signal with Gaussian fit, definition of peak height and jitter, c) flattened images of the peak height, TOF-jitter and distance, extracted from one lidar scan, and a combination of jitter- and peak height-image via summation in fourier space, d) visualization of the data in 3D via distance information, and color mapping with corresponding peak height and jitter data}
\end{figure*} 
We further use SNSPD1 for scanning centimeter-sized objects representative of the streets of Münster, Germany, as shown in Fig.$\,$\ref{fig:image} a), for obtaining "full waveform" data. The distribution of the TOF-detector output signals, exemplary shown in Fig.$\,$\ref{fig:image} b), allows for a simple extraction of temporal and intensity information. The peak height and the TOF-jitter can yield information about the target surface. Fig.$\,$\ref{fig:image} c) displays flattened visual representations of the peak height and TOF-jitter of each pixel in the field of view.
The peak height data reveals information connected to the reflectivity of the materials, e.g. comparing the metal and rubber of the bike wheel. Moreover, topological differences are apparent in the peak height image, as surfaces with large angles to the beam show lower peak heights and edges of objects show a large drop in peak height. This can reveal topological information, e.g. spokes of the bike, even for features exceeding the lateral resolution of the system, here given by the beam spot size. This is a common imaging effect limited by the convolution of the beam spot size and the feature size of the scanned object, which can also be described with the point spread function of the imaging system.
The jitter data shows a similar but inverted behavior, as surfaces with large angles to the laser beam and edges of objects show increased jitter. The edge effects occur due to measuring $t_{TOF}$ over the spatial extent of the beam spot, which integrates reflections from an area, then represented as a point or pixel in the dataset.
As the jitter and peak height show inverse, but topologically sensitive behavior, we perform a data fusion technique, by combining the information on the "full waveform" data into a single image. We reduce noise with a common median image filter, normalize the data and add the images in Fourier space. Converting back to real space then yields an image of combined height and jitter, depicted in Fig.$\,$\ref{fig:image} c). Further, we map this surface response data onto the 3D spatial data to arrive at a final image of the data fusion technique, depicted in Fig.$\,$\ref{fig:image} d). For the visualization in 3D, we utilize a common computer graphics algorithm, a delaunay triangulation, to generate a polygon surface from the point cloud. We color the polygon with the surface response data of the combined peak height and jitter. Some gaps in the polygon occur because the measurement points in space are not spaced equidistantly, due to scanning with steps of constant angles.

\subsection{\label{sec:results_improved} Improvement on range resolution}
Superconducting nanowire detectors are primarily used in single-photon counting applications and consequently SNSPD jitter and range resolution in corresponding lidar systems are often explored in the regime of very low photon flux. However, the jitter behavior at high photon flux is not fully explored in this context. Here, we study the dependence of the TOF-jitter on the average photon number per pulse arriving at the SNSPD. We control the number of photons reflected off the target with a VOA in the ranging path and by operating the EDFA in a regime that avoids pulse broadening. We here utilize a reflective target at a distance of a few mm to increase the range over which the average photon number can be varied in TOF measurements. We determine the average number of photons per pulse arriving at the SNSPD, $\bar n$, from monitoring the average power of the laser pulses at the VOA before they are transmitted into the free space section. We here exploit the higher absorption efficiency of SNSPD2, which facilitates the simultaneous absorption of multiple incident photons in the nanowire, thus resulting in a multiphoton detector response. As compared to conventional meander-shaped SNSPDs that are designed for absorbing photons from a ten's of square micrometer-sized optical fiber mode under normal incidence, our implementation exploits the fact that waveguide-integrated SNSPD allow for achieving high absorption efficiency in short nanowires, which benefit high timing accuracy.  
\begin{figure}
\includegraphics{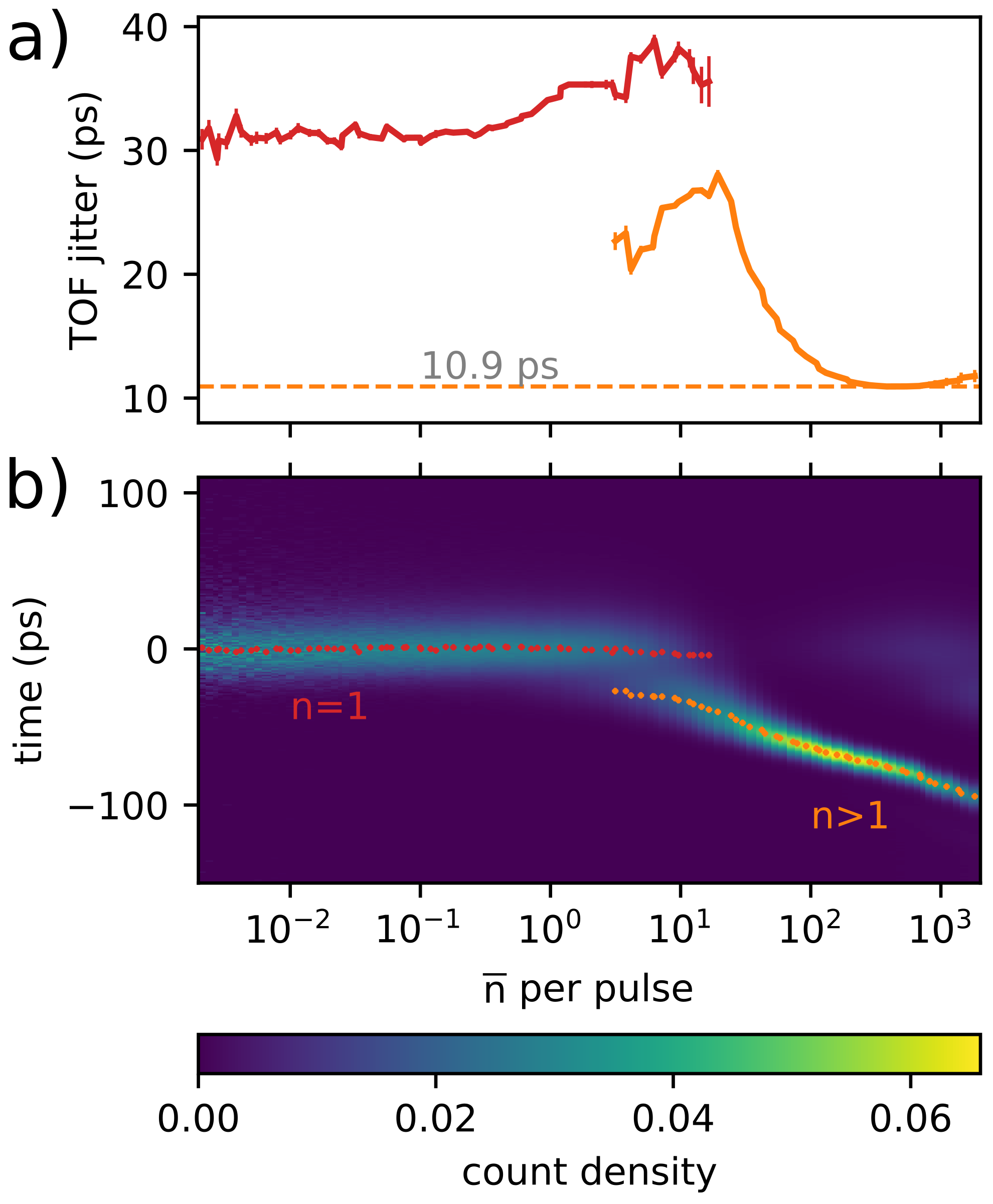}
\caption{\label{fig:jitter} TOF-measurements for varying average photon numbers $\bar n$ per pulse arriving at the SNSPD, Gaussians fitted for $n=1$ (red) and $n>1$ (orange) detections, a) FWHM TOF jitter values, b) TOF histograms with latency time on y-axis, $\bar n$ on the x-axis and the count density in color code, dots showing center of gaussian fits}
\end{figure}
Fig.$\,$\ref{fig:jitter} a) shows the dependence of the TOF-jitter on $\bar n$ and TOF-histograms of the count density for observable latency times are shown in Fig.$\,$\ref{fig:jitter} b). The count density was calculated by normalizing the histogram for each $\bar n$ individually.
For low photon numbers ($\bar{n}<<1$), only single photon absorption events in the nanowire, n=1, produce detector output signals of constant TOF jitter and latency time. 
As the average photon number in a pulse increases, the absorption of multiple photons ($n>1$) in the nanowire becomes more probable, then producing multiphoton detector output signals. These multi-photon detections exhibit shorter latency and rise times in the voltage response of SNSPDs, compared to single-photon detection events, leading to an earlier detection event and reduced TOF-jitter. This behavior is consistent with intrinsic photon number resolving SNSPDs, where higher kinetic inductance even allows for separating the signals corresponding to the absorption of $n$ different photon numbers \cite{Jaha2024}. In our lower kinetic inductance devices it remains possible to distinguish $n=1$ single photon contributions but the $n>1$ absorption events produce detector responses with a continuous reduction of their latency times as $\bar{n}$ increases, as shown in Fig.$\,$\ref{fig:jitter} b). 
We find the smallest jitter of 11$\,$ps at $\bar n \approx 545$, which corresponds to a single-shot resolution of 1.65$\,$mm. 

\section{\label{sec:disc} Conclusion and Discussion}
We have reported a TOF lidar system that leverages the benefits of waveguide-integrated superconducting nanowire single photon detectors, combining excellent temporal photon counting performance with high photon absorption efficiencies, for achieving high resolution ranging. 
For multi-shot measurements, averaging multiple measurements per pixel, we find a sub-mm range resolution of 750$\,\mu$m for scanning an aluminum plate at a distance of 51$\,$cm at 1550$\,$nm wavelength. Our results compare favorable to state-of-the-art TOF-lidar-measurements of a single point with 0.8$\,$mm resolution \cite{Staffas2024}.
For single-shot measurements, i.e. a single detection per pixel, the range resolution heavily depends on the temporal jitter of the detector. We fabricated a waveguide integrated SNSPD (design SNSPD1) with a jitter of 21$\,$ps and measured an average of (27.6$\pm$3.3)$\,$ps TOF-jitter for the same target and distance, corresponding to a range resolution of 4.1$\,$mm. \\
"Full waveform" images recorded with our lidar system indicate sections of higher temporal spread revealed by the low jitter of the SNSPD of our photon counting lidar, with a particular sensitivity for edges. Combining the reflectivity, jitter and spatial data, the "full waveform" images hold great potential for automated generation of digital twins that can be used as input for improved training routines of machine vision or artificial intelligence (AI) applications for object recognition or material distinction. Depending on the computational cost versus the dimension of data input, the three quantities, reflectivity, jitter and spatial data, can be utilized in such applications in combined or separate form.\\
Moreover, we present a reduction of the TOF-jitter by operating SNSPDs in the multiphoton detection regime. For SNSPD2 we find an improvement in timing accuracy from 31$\,$ps to 11$\,$ps, corresponding to 1.6$\,$mm range resolution. One effect reducing the jitter can be attributed to the energy dependent latency response of the SNSPDs \cite{Korzh2020,Allmaras2019}. An additional mechanism of this jitter reduction is described by the statistical elimination of the Fano-fluctuation and minimization of geometrical contributions due to large numbers of photons being absorbed at different locations of the wire \cite{Sidorova2018}, creating interdependent hotspots in the superconducting order parameter. Further experiments investigating the dependence of jitter performance on photon flux for very thin and short nanowires, limiting the effect of hotspot interactions, may provide insight into the physics of the jitter reduction.\\
Implementing the multiphoton resolution improvement in a lidar setting can be accomplished by measuring the rise time of the detection signals simultaneously to the TOF. While the higher n photon detections exhibit a lower temporal jitter, they are shifted in latency due to the different shape of the detection signal. Knowledge of this shape or rise time allows for a correction of the latency and TOF. This can be measured by splitting the detection signal and measuring the time of incidence at an additional channel of the TDC, but at a different trigger level voltage.\\
A further benefit of waveguide-integrated SNSPDs over conventional fiber-coupled meander-type nanowire solutions for lidar applications is the possibility of integrating nanophotonic functionalities within the optical channels. Phase shifters and modulators, for example, provide important on-chip scanning solutions for beam steering or optical phased arrays and have already been shown to be compatible with SNSPDs \cite{Grottke2021,Lomonte2021}. Exploiting also the favorable scaling prospects of waveguide-integrated SNSPDs \cite{Haeussler2023} highly integrated high-resolution lidar systems become possible with exciting possibilities in the generation of digital twins and remote object recognition.

\section{Acknowledgments}
We acknowledge support from the European Union’s Horizon 2020 Research and Innovation Action under Grant Agreement No. 899824 (FET-OPEN, SURQUID), and from its successor program Horizon Europe under Grant agreement No. 101135288 (EPIQUE). The authors further acknowledge the financial support from the Federal Ministry of Education and Research of Germany in the frameworks of project MultiQomm (Grant No. BMBF 16KIS1800) and of project SeQuRe (Grant No. BMBF 16KIS2149). We thank Simone Ferrari for helpful discussions and the Münster Nanofabrication Facility (MNF) for their support in nanofabrication-related matters.

\section{References}
\bibliography{bib_SNSPDLidar_jitter}

\end{document}